\documentclass{aa}  
\usepackage{graphicx}
\usepackage{amsmath}
\usepackage{natbib}
\bibpunct{(}{)}{;}{a}{}{,}
\usepackage{txfonts}
\def\kms{km~s$^{-1}$}

\def\sU{\sigma_{\rm U}}
\def\sU'{\sigma_{\rm U'}}
\def\sV{\sigma_{\rm V}}
\def\sV'{\sigma_{\rm V'}}

\begin{document}
\titlerunning{Age heterogeneity of moving groups}
  \title{On the age heterogeneity of the \\ Pleiades, Hyades and Sirius moving groups}   

   \author{B. Famaey
          \inst{1}
          \and
          A. Siebert
          \inst{2,3}
          \and
          A. Jorissen
          \inst{1}
          }

  \offprints{B. Famaey}

   \institute{Institut d'Astronomie et d'Astrophysique, Universit\'e
Libre de
  Bruxelles, CP 226, Boulevard du Triomphe, B-1050 Bruxelles, Belgium            
         \and
Astrophysikalisches Institut Potsdam, An der Sterwarte 16, D-14482 
Potsdam, Germany
         \and
Observatoire de Strasbourg, 11 Rue de l'Universit\'e, F-67000 Strasbourg, France
  }

   \date{Received ...; accepted ...}

 
  \abstract 
   {}
   {We investigate the nature of the classical low-velocity structures in the local velocity field, i.e. the Pleiades, Hyades and Sirius moving groups. After using a wavelet transform to locate them in velocity space, we study their relation with the open clusters kinematically associated with them.
}
   {By directly comparing  the location of moving group stars in parallax space to the isochrones of the embedded clusters, we check whether, within the observational errors on the parallax, all moving group stars could originate from the on-going evaporation of the associated cluster.
}
   {We conclude that, in each moving group, the fraction of stars making up the velocity-space overdensity superimposed on the background is higher than the fraction of stars compatible with the isochrone of the associated cluster. These observations thus favour a dynamical (resonant) origin for the Pleiades, Hyades and Sirius moving groups.
} 
   {}


   \maketitle
%

\section{Introduction}

It has been known for  a long  time that the  local velocity  field in  the solar neighbourhood is  clumpy, and that most  of the observed clumps  are made of spatially $unbound$ groups of stars,  called moving groups (e.g., Eggen 1958, 1960, 1975, 1983). 

Although unbound, most of these  moving  groups  however  share  the  kinematics  of  well-known  open clusters.  The  best  documented   low-velocity  groups  (e.g., Famaey et  al. 2005; Chereul  et al.  1998, 1999;  Dehnen 1998; Montes et  al. 2001; Ecuvillon et al. 2007)  are the Hyades  moving group (e.g., Famaey  et al.  2007) associated  with the Hyades  cluster  (600~Myr)  and   the  Sirius  moving group  associated  with  the evaporating UMa  star cluster (300~Myr). Another kinematic  group called the Local Association  or Pleiades  moving group is a  set of stars mostly associated with the Pleiades cluster (100~Myr), and with a few other young clusters with ages all below 150~Myr ($\alpha$ Persei, NGC~2516, IC~2602 and Scorpio-Centaurus). Finally, the  large group consisting of stars  lagging behind the galactic rotation and moving outward  in the disk (e.g., Blaauw 1970, Raboud et al. 1998, Famaey et al. 2005, Bensby et al. 2007) is called the Hercules stream.

The classical hypothesis of Eggen is that  those low-velocity moving groups are in fact remnants of clusters  which partly evaporated with time. Indeed, their prominence close to the Sun's position in velocity space without any vertical flow (e.g., Seabroke et al. 2007) is not compatible with the hypothesis that they are the result of merger events with satellite galaxies. Such satellite debris have been identified in the solar neigbourhood (Helmi et al. 2006, Arifyanto \& Fuchs 2006, Dettbarn et al. 2007), but containing much less stars and moving with higher velocities. However, Eggen's hypothesis is still debatable because at least some of the moving groups may also be  generated   by resonant mechanisms linked with the non-axisymmetry and non-stationarity of the Galaxy. E.g., a  rotating $bar$ at the centre of  the Milky Way is now thought to be able to create the Hercules stream if the Sun is located near the  outer  Lindblad resonance (OLR) of the bar. The existence of such a kinematical group could be due to the  coexistence at the OLR of  orbits elongated along and  perpendicular to the bar's major axis (Dehnen 2000), or to an overcrowding of chaotic regions in velocity space induced by the bar near the OLR (Fux 2001). On the other hand, moving groups with lower velocities such as the Hyades and the Pleiades could still be linked with the $spirality$ of  the Galaxy. E.g., a  series of strong  transient spiral arms with  their mean  corotation at  the solar  galactocentric radius  have been shown to  produce similar structures in the  local velocity distribution (De  Simone et  al. 2004), while a  group such  as the  Hyades  could also correspond  to nearly-closed  orbits  trapped at  the  $4:1$ inner  Lindblad resonance (ILR) of a two-armed spiral density wave (Quillen \& Minchev 2005).

Eggen's  hypothesis  and  the  dynamical  perturbations by the bar and the spiral arms are  of  course  not mutually incompatible scenarii. Clusters do evaporate over time and, at some intermediate stage, an  unbound group of stars with  similar velocities must appear (Woolley  1961). On  the other hand,  large spiral  perturbations are known to indeed exist in the  Galaxy, and  must have an effect  on local stellar  kinematics.  So,  for  an  individual  moving  group,  how  can  we observationally distinguish between a  pure Eggen scenario, a pure dynamical scenario and a mix of both?

Famaey et al. (2005) used a maximum-likelihood method (Luri et al. 1996) to model a sample of K and M giant stars (combining Hipparcos parallaxes with Tycho-2 proper motions and CORAVEL radial velocities) in order to assign each star to a kinematic base group. This allowed to identify three main moving groups (Hercules, Sirius and Hyades-Pleiades) and to plot the corresponding stars in Hertzsprung-Russell (HR) diagrams: the similarity between the HR diagrams of stars in the moving groups and in the background was noted, which argued in favour of a dynamical (resonant) origin for the groups rather than a coeval (evaporated cluster) origin. However, this method, although extremely powerful in order to yield unbiased distance estimates, presented two shortcomings: first, the method revealed itself unable to separate distinct substructures on a scale smaller than 30~km/s (e.g., to separate the Hyades from the Pleiades moving group); second, although the similarity of HR diagrams for the moving groups and the background was striking, it did not statistically exclude the possibility of such HR diagrams being the result of coeval groups superimposed on the background. However, a complementary analysis of the Hyades moving group, by Famaey et al. (2007), based on the Geneva-Copenhagen catalogue of F and G dwarfs (Nordstr\"om et al. 2004), revealed that the number of stars in the group with respect to the total number of stars in the background was independent of mass, contrary to the expectations based on the difference between the initial mass function (IMF) of an evaporating cluster and the present-day mass function (PDMF) of the background, arguing again in favour of a resonant origin for the Hyades moving group.

Here, we reanalyze the data of Famaey et al.~(2005) in order to circumvent the two shortcomings of the maximum-likelihood method. First, we use a wavelet transform (see Chereul et al. 1998, 1999 who had not used the radial velocities that are now available) in order to identify the structures on scales smaller than 30~km/s. Second, by directly comparing  the location of moving group stars in parallax space to isochrones  of the  embedded clusters, we independently confirm the dynamical nature  of the groups (Famaey et al. 2005, 2007) in a robust statistical way.

\section{Data}

Fifteen years of observations with the CORAVEL spectrometer on the Swiss 1-m telescope at the {\it Observatoire de Haute Provence} yielded the radial velocities of Hipparcos stars later than about F5 in the northern hemisphere. This unique database of radial velocities measured with a typical precision of 0.3~km/s, combines a  high precision and the absence of kinematic bias. The details of the sample of K and M giants used in this paper (combining Hipparcos parallaxes with Tycho-2 proper motions and CORAVEL radial velocities) can be found in Famaey et al.\ (2005).

\section{Locating the moving groups}
\label{Sect:wavelet}

The parametric maximum-likelihood technique of Luri et al. (1996) and Famaey et al. (2005)  is not suited to find small structures  in the velocity space and cannot separate for example the Hyades and Pleiades moving groups.  To detect these small-scale structures, we use  a wavelet transform technique on the 2-D velocity field defined by the $U$ and
$V$  velocities ($U$ is the velocity towards the galactic center, $V$ the velocity in the direction of Galactic rotation, both with respect to the Sun). The  method  is similar  to  Chereul et  al.  (1998,  1999)
analysis but for clarity, the main steps are summarized below. 

Let us however note that the method used in Famaey et al. (2005) had the great advantage of giving an unbiased statistical estimate of the distance of individual stars, unlike the inverse of the parallax affected by the Lutz-Kelker bias. In the process of analyzing the distribution of stars in velocity space with the wavelet transform, we thus have to make use of the results of Famaey et al. (2005) concerning the distances of individual stars.

The  wavelet  transform by  an  oscillating  and  zero-integral {\em  mother
wavelet} $\Psi(x)$ of a real one-dimensional signal $F(x)$ is defined as:\\
\begin{equation}
W_s(i)=\frac{1}{\sqrt{s}}\int^{+\infty}_{-\infty}F(x)\cdot \Psi^{\ast}\left(\frac{x-i}{s} \right) {\rm d}x\\
\end{equation} 
where $s$  is the  scale and $i$  the position  of the analysis.  The values
$W_s(i)$  are called  the wavelet  coefficients. The  original signal $F(x)$  can then recovered by a double summation of $W_s(i)\cdot\Psi(\frac{x-i}{s})$ over the variables $s$ and $i$.

Following Chereul et  al. (1998, 1999), we use the {\it \`a  trou} algorithm (e.g., Holschneider et al. 1989)  to  compute  the  wavelet decomposition  of our  2-D velocity  field. 
This algorithm relies  on the definition  of  the  mother  wavelet  $\Psi(x)$ as  the  difference  at  two different  scales  of the  same  smoothing  function  (or scaling  function) $\Phi(x)$:

\begin{equation}
\Psi(x)=\Phi(x)-\frac{1}{2}\Phi(\frac{x}{2})\, ,
\end{equation}

\noindent and on the relation

\begin{equation}
\frac{1}{2^{s+1}}\Phi(\frac{x}{2^{s+1}})=\sum_{l=-2}^{2}h(l) \Phi(\frac{x}{2^s}-l)\, .
\end{equation}

Here  $h(l)$   is  a  one-dimentional  discrete  low   pass  filter  applied
iteratively to compute the smoothed  distributions $C_s$ at scale $s$. As in
Chereul et al. (1998, 1999), we use 
$h(l)=\{\frac{1}{16},\frac{1}{4},\frac{3}{8},\frac{1}{4},\frac{1}{16}\}$ for $l=\{-2, -1, 0, 1, 2\}$, and
$\Phi(x)=(|x-2|^3-4|x-1|^3+6|x|^3-4|x+1|^3+|x+2|^3)/12$. 

For our 2-D analysis, we define $\Phi(x,y)=\Phi(x)\cdot\Phi(y)$. 
The coefficients $C_s(i,j)$  [i.e. the smoothed signal at scale $s$ and pixel $(i,j)$] are
given by the recursive relation
\begin{equation}
C_s(i,j)=\sum_{l=-2}^{l=2}\sum_{m=-2}^{m=2}h(l)h(m)\,
C_{s-1}(i+2^{s-1}l,j+2^{s-1}m)\, ,
\end{equation}
$C_0$  being the  observed distribution  $F(U,V)$ convolved  by  the scaling
function $\Phi(x,y)$ (see Chereul et  al 1999, Fig.~1 for a detailed schematic
overview of  the procedure).  With  this relation, the distance  between two
bins increases by a factor 2  between the scales $s$ and $s+1$.  The wavelet
coefficients $W_s(i,j)$, are then given by the signal difference between two
scales

\begin{equation}
W_s(i,j)=C_{s-1}(i,j)-C_s(i,j)\, .
\end{equation}

It is clear  from this relation that the wavelet  coefficients at each scale
have  the  same spatial  coverage  in the  velocity  space  as the  original
distribution, which permit to localize the structures in the velocity space.
In  practice, we bin  the 2-D  velocity field  $F(U,V)$ on  a 500$\times$500
squared grid from -250~km/s to +250~km/s. We then analyze the observed histogram
on 6 dyadic scales $s$ corresponding to 5, 7, 10, 14, 20 and 28~km/s with the
previous definition.\\

The original distribution is subject to Poisson noise and noise fluctuations
induce  non-zero   wavelet  coefficients  at  each   scale.  These  non-zero
coefficients generate artifacts mimicking substructures and must be removed
prior to the scientific analysis.
To  reject  the  noise-induced  coefficients  and select  only  the  regions
corresponding  to  real  overdensities,  we  again  follow  Chereul  et  al.
(1999). We perform  numerical simulations of a 2-D  Poisson noise and follow
rigourously the  same procedure (binning and wavelet  transform) to estimate
the amplitude  of the  wavelet coefficients generated  by the noise  at each
scale.
We then compute a treshold $\lambda_s^+(i,j)$ at each scale and each pixel such that the probability that  the  noise wavelet coefficient is larger than this threshold is $P=10^{-4}$.
We subsequently apply a hard thresholding  to clean the wavelet
coefficients using

\begin{eqnarray}
W_s(i,j)=&
\begin{cases}
W_s(i,j)\,\,\textrm{ if}\,\, W_s(i,j) \geq \lambda_s^+(i,j)\\
0 \,\, \textrm{ otherwise}
\end{cases}
&\, ,
\end{eqnarray}


The small-scale structure appears then very clearly on the 14~km/s scale, which is plotted on Fig.~1.

\begin{figure}
\centering
\includegraphics[width=7cm]{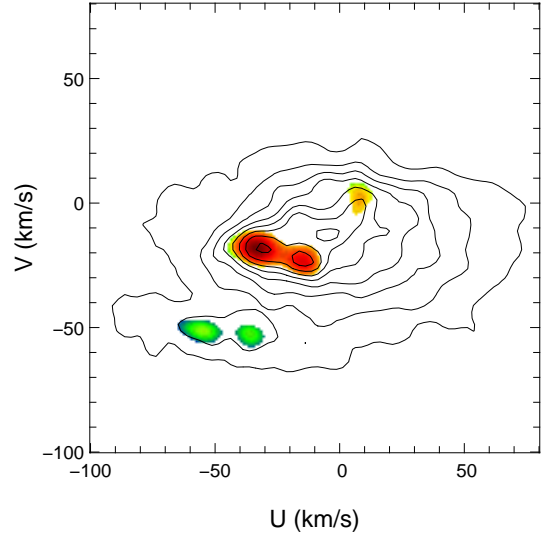}
\caption{Contours and wavelet coefficients in the [$U,V$] plane for the Famaey et al. (2005) sample. Black: isocontours
  of the velocity distribution smoothed over 14~km/s.  The color scales with
  the  wavelet coefficients at  scale 4  (14~km/s) after  thresholding.  The four structures observed  in the wavelet space correspond  to the Hercules stream  (bottom green structures),  Pleiades  and Hyades  (middle red structures from right  to left) and Sirius (yellow, top).  The limit of the structures in the wavelet space are used to select the samples for further analysis.}
\label{Fig_UVwavelet}
\end{figure}

The wavelet coefficients at scale 4 (see Fig.~1) are non-zero in the three boxes of the [$U,V$]-plane corresponding to the Pleiades ($U\simeq[-23,-9]$, $V\simeq[-28,-18]$), the Hyades ($U\simeq[-41,-23]$, $V\simeq[-24,-12]$), and Sirius ($U\simeq[0,10]$, $V\simeq[-5,8]$). Interestingly the Hercules stream is also recovered, but is separated in two different substructures ($U\simeq[-65,-49]$ and $[-40 -30]$, $V\simeq[-55,-47]$), a fact which had not been identified so far in previous studies.

The most interesting result from this wavelet analysis is thus the similarity between the small-scale structure of the Hyades-Pleiades moving group and the Hercules stream. They both appear as a {\it pair} of groups. If they are both of dynamical origin, this could provide an important new constraint on dynamical models designed to reproduce them (see e.g. figure 9 of De Simone et al. 2004). 

\section{Age of the moving groups}

Now that the locations of the substructures have been precisely identified in velocity space, we look whether these subtsructures result or not from the sole evaporation of the main clusters associated with them. Since the dynamical nature of the Hercules stream is well-proven from a detailed analysis of its chemical heterogeneity (Bensby et al. 2007), we concentrate hereafter on the Pleiades, Hyades and Sirius moving groups. 

For each zone in velocity-space determined  hereabove with the wavelet technique, we shall first estimate the fraction of stars in the moving group and in the background. Then, we shall assume that all the stars of the group are coeval and thus located on a single isochrone in the Hertzsprung-Russell diagram. We can then compute for each star the parallax that would correspond to this isochrone. Then, the relative difference (normalized by the error) of the Hipparcos parallaxes and isochrone ones in parallax space should follow a gaussian of mean 0 and standard dispersion 1 after the distribution of background stars in parallax space has been subtracted. If this is not the case, the moving group will be proven to be, at least partially, of dynamical (resonant) origin.

\subsection{Estimating the fraction of stars in the moving groups}

\begin{figure}
\centering
\includegraphics[width=7cm]{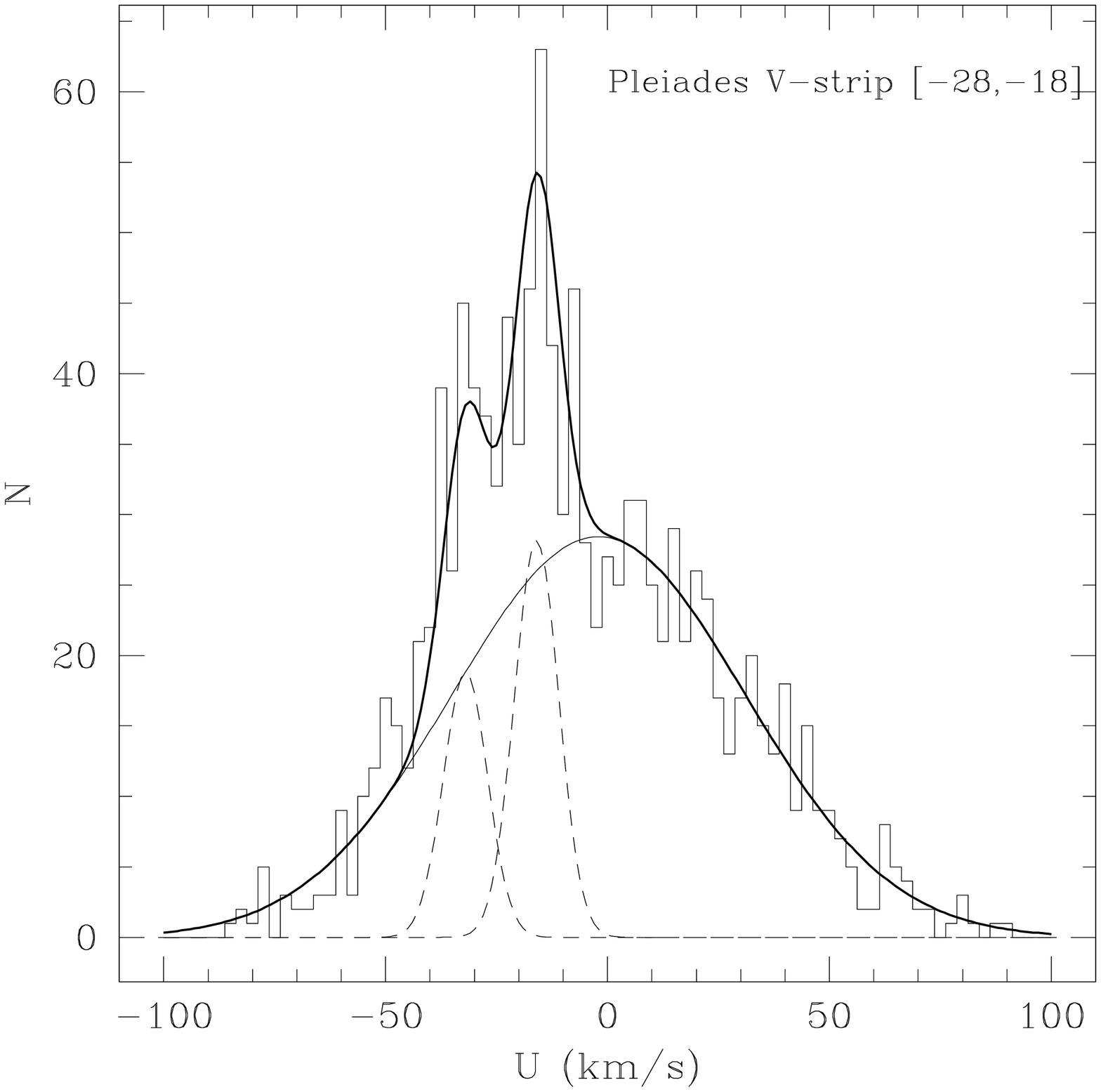}
\includegraphics[width=7cm]{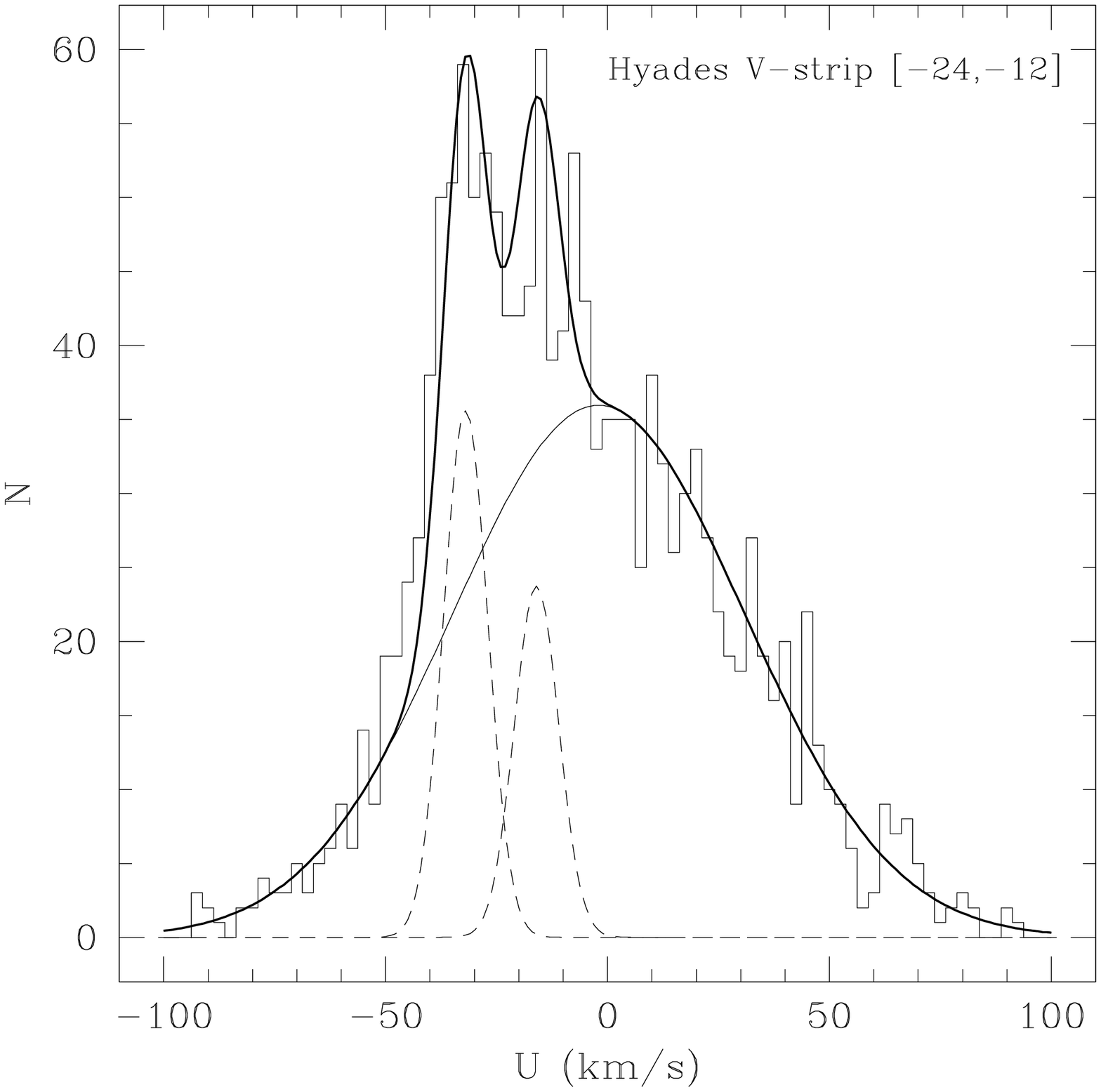}
\mbox{}\vspace{-5mm}
\includegraphics[width=7.7cm]{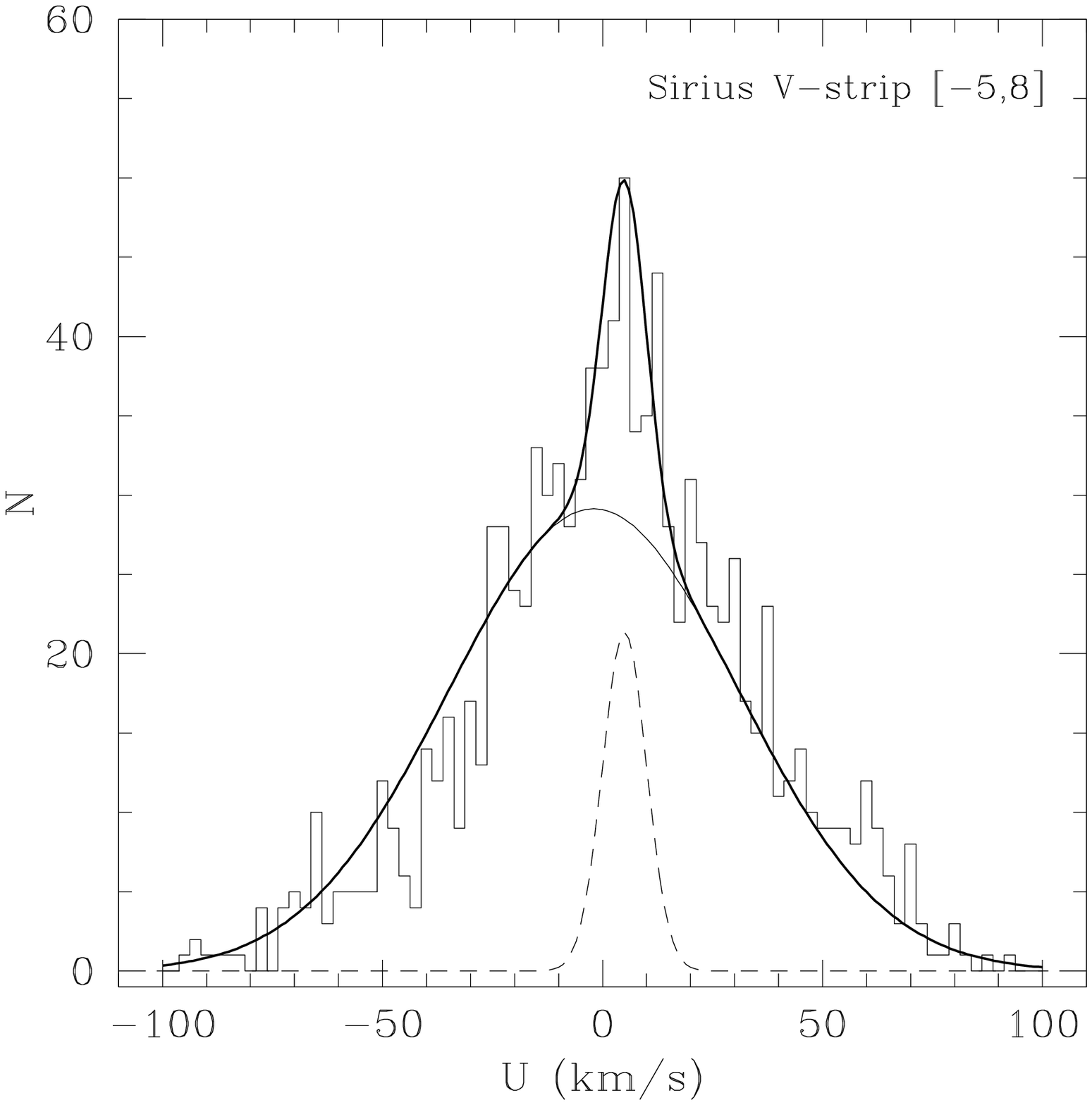}
\caption{{\tiny Top: $U$ distribution of stars from the Famaey et al. (2005) sample for the Pleiades strip $-28 \le V$~(\kms)~$\le -18$. The thick solid
  line is a fit to the observed distribution, resulting from the
  superposition of three gaussians (with parameters listed in
  Table~\ref{Tab:gaussian_ple}),
  corresponding to the background  component (thin solid line), to
  the full Pleiades component and partial Hyades component (thin dashed lines).
Middle: same as top for the Hyades strip $-24 \le V$~(\kms)~$\le -12$ (with parameters listed in Table~\ref{Tab:gaussian_hy}). Bottom: same as top and middle for the Sirius strip $-5 \le V$~(\kms)~$\le 8$ (with parameters listed in Table~\ref{Tab:gaussian_si})}
}
\label{Fig:histoband}
\end{figure}

In order to separate the background from the Pleiades
component, the histogram of stars in the strip $-28 \le V$~(\kms)~$\le -18$ (corresponding to
the range spanned by the Pleiades in Fig.~\ref{Fig_UVwavelet}) was plotted in Fig.~\ref{Fig:histoband},
and three separate gaussians fitted to the total histogram (see Famaey et al. 2007): one
representing the background, another representing the Pleiades and the
last one {\it partially} representing the Hyades (which partially contaminates the $U$-distribution in this strip). Since the Hyades $V$-strip is slightly different from the Pleiades one, the operation has been repeated for the Hyades with the strip $-24 \le V$~(\kms)~$\le -12$. Finally, for Sirius, only two gaussians were fitted to the histogram in the strip $-5 \le V$~(\kms)~$\le 8$. The parameters of the gaussians yielding the best fit to the data are
given in Table~\ref{Tab:gaussian_ple} for the Pleiades strip, Table~\ref{Tab:gaussian_hy} for the Hyades strip, and Table~\ref{Tab:gaussian_si} for the Sirius strip. It is worth noting that the
$\sigma$ of the background gaussian (33~km/s) is consistent with the
result listed in table~2 of Famaey et al.~(2005), and that the
parameters of the Hyades and Pleiades gaussians are precisely identical (except of course for $\bar{U}$ and for the absolute number of stars present in the strip).  

\begin{table}[hbtp]
\caption{Parameters of the gaussians fitting the $U$ distribution of
  stars from Famaey et al. (2005), in the $V$-strip
  corresponding to the Pleiades, $ -28 \le V$ (\kms) $\le -18$. The parameters $k_i$ give the relative fractions $N_i/N_{\rm
  tot}$, where $N_i$ is the total number of stars in 
  component $i$, and $N_{\rm tot}=1176$ is the total number of stars in the
  $V$-strip. The fraction $f_s$ is the
  relative fraction of stars from the Pleiades moving group with respect to
  {\it all} stars in the restricted $\pm 1.4 \sigma_{U}$ range
  around the gaussian average, corresponding to the Pleiades zone where the wavelet coefficients are non-zero at scale 4 (see Sect.~3).}
\label{Tab:gaussian_ple}
\begin{tabular}{llllllll}
                  & background & Pleiades\\
$\bar{U}$ (\kms)  & $-2$ & $-16$ \\
$\sigma_U$ (\kms) & 33   & 5     \\
$k$               & 0.8  & 0.12  \\
$f_s$           &    & 0.46  \\ 
\end{tabular}
\end{table}

\begin{table}[hbtp]
\caption{Same as Table~1 in the $V$-strip
  corresponding to the Hyades, $ -24 \le V$ (\kms) $\le -12$, with $N_{\rm tot} = 1487$.}
\label{Tab:gaussian_hy}
\begin{tabular}{llllllll}
                  & background & Hyades       \\
$\bar{U}$ (\kms)  & $-2$ & $-32$  \\
$\sigma_U$ (\kms) & 33   & 5      \\
$k$               & 0.8  & 0.12   \\
$f_s$            &     & 0.52  
\end{tabular}
\end{table}

\begin{table}[hbtp]
\caption{Same as Table~1 in the $V$-strip
  corresponding to the Sirius moving group, $ -5 \le V$ (\kms) $\le 8$, with $N_{\rm tot} = 1071$.}
\label{Tab:gaussian_si}
\begin{tabular}{llllllll}
                  & background & Sirius       \\
$\bar{U}$ (\kms)  & $-2$ & $5$  \\
$\sigma_U$ (\kms) & 33   & 5      \\
$k$               & 0.9  & 0.1   \\
$f_s$             &     & 0.35  
\end{tabular}
\end{table}

The relative contribution of each moving group with
respect to the background at their corresponding location in the
  [$U,V$]-plane is then obtained by integrating these
gaussians over a range in $U$ corresponding to $\pm 1.4\sigma$. This has been chosen to match the regions where the wavelet coefficients are non-zero (precisely in the case of the Pleiades, approximately in the case of the Hyades and Sirius, see Fig.~\ref{Fig_UVwavelet}), i.e. $U \in [-23,-9]$ for the Pleiades, $[-39,-25]$ for the Hyades, and $[-2,12]$ for Sirius. We find that, in these regions of velocity space, the percentages of stars physically belonging to the Pleiades, Hyades and Sirius moving groups are respectively 46\% (111/239 stars in the Pleiades), 52\% (150/290 stars in the Hyades) and 35\% (60/169 in Sirius).

\subsection{Isochrone analysis in parallax space}

\begin{figure}
\centering
\includegraphics[width=7cm]{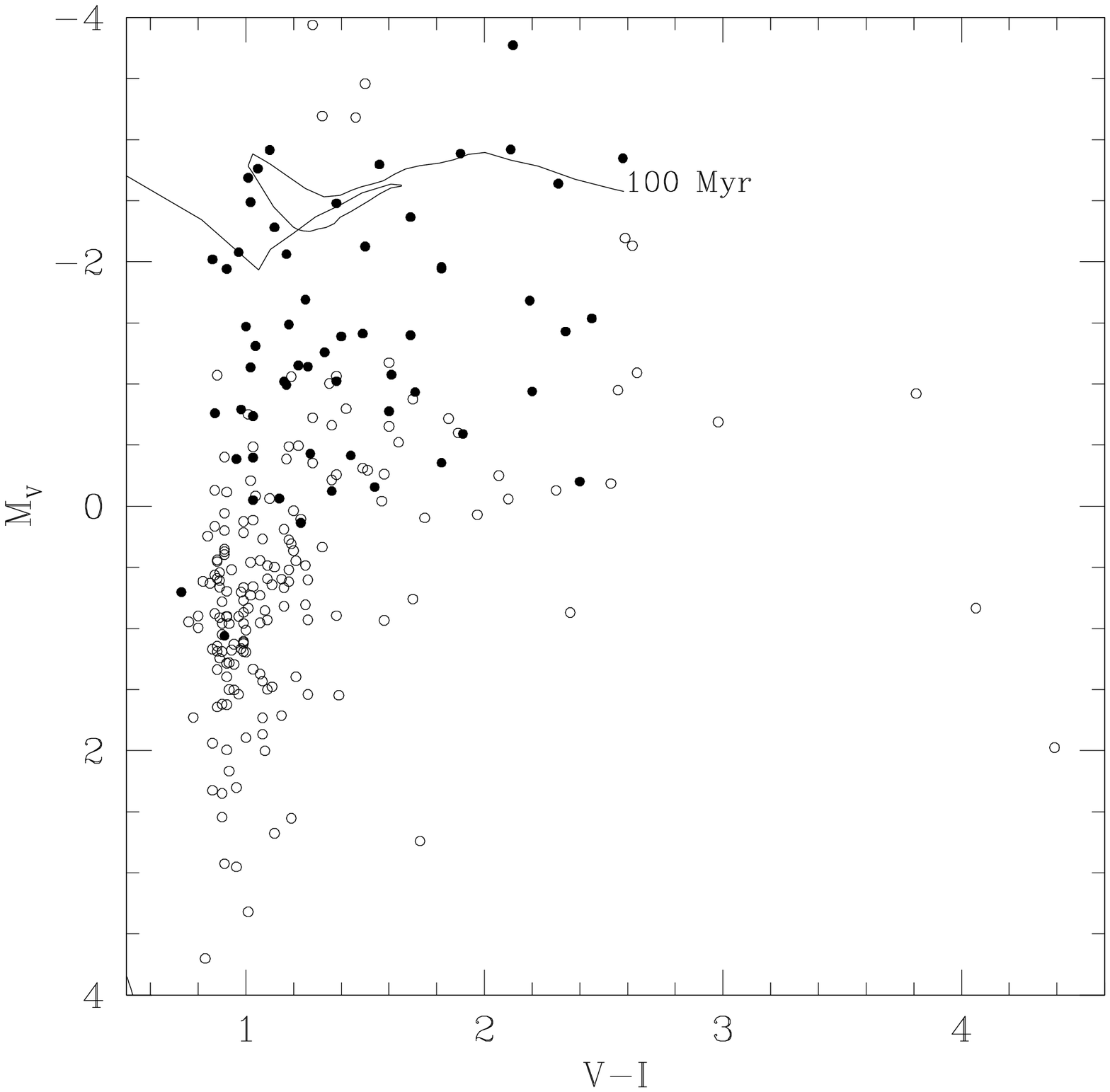}
\includegraphics[width=6.5cm]{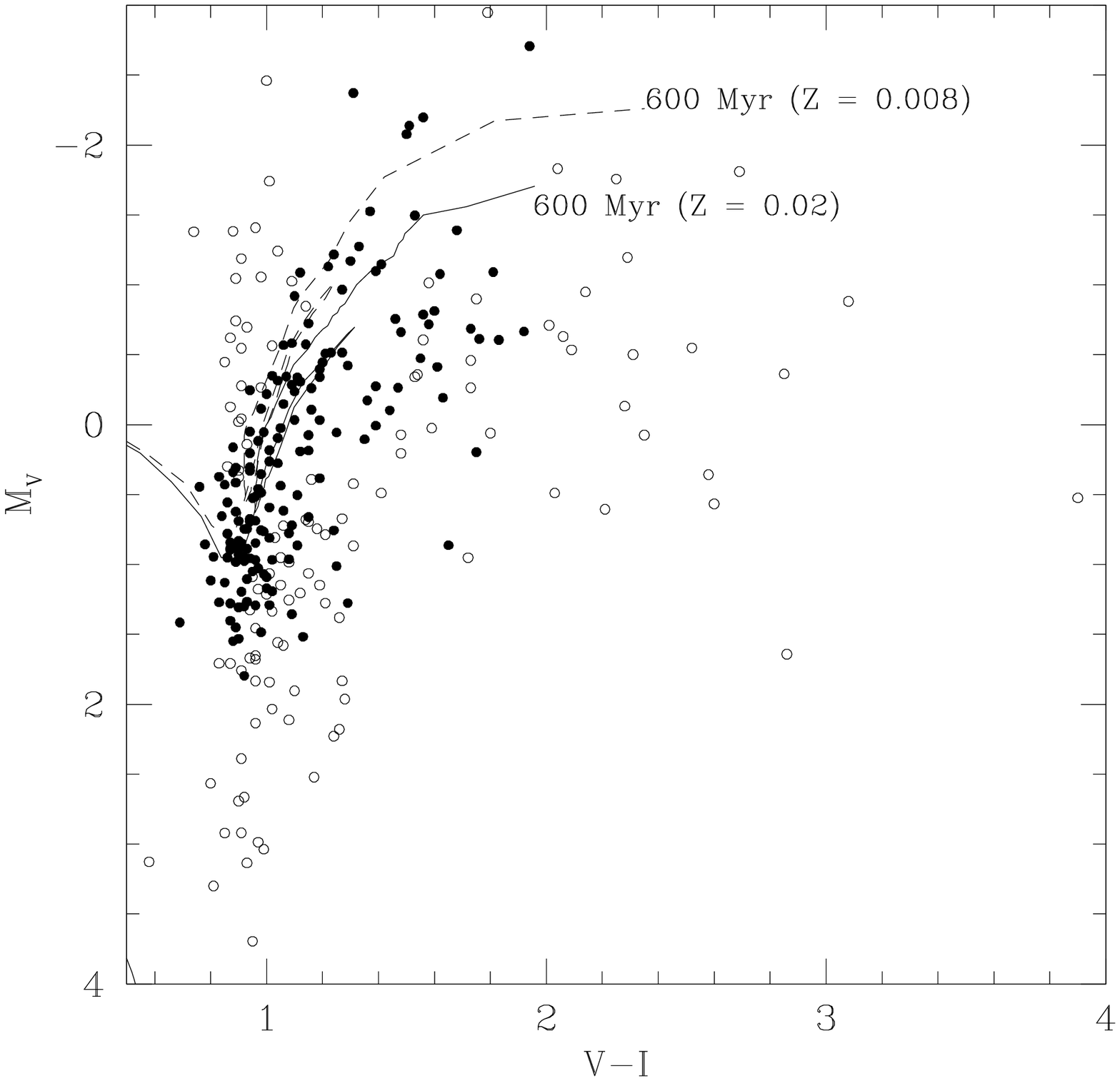}
\mbox{}\vspace{-3mm}
\includegraphics[width=7cm]{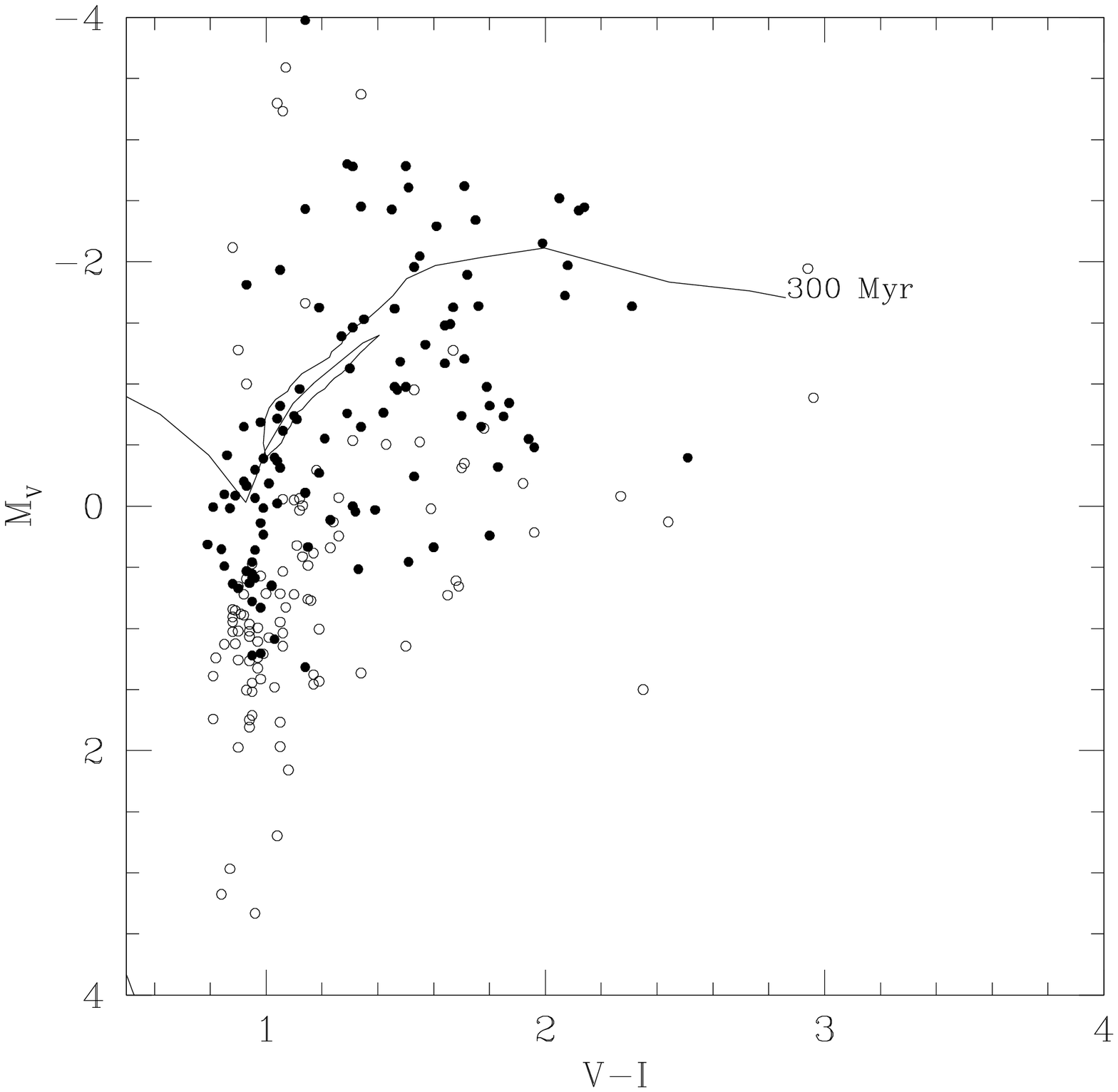}
\caption{{\tiny Top: Comparison of the 100~Myr isochrone ($Z=0.02$) from Schaller et al.~(1992) and
Lejeune \& Schaerer~(2001) with the stars kinematically
associated with the Pleiades cluster/moving group. Note that 
absolute magnitudes have been derived from a simple inversion of the parallax, and are thus biased, which is not a problem for the subsequent analysis done in parallax space. Stars whose parallax is within $2\sigma$ of the isochrone parallax are marked with filled circles while others are marked with open circles. Middle: Same for the stars kinematically
associated with the Hyades cluster/moving group and the 600~Myr isochrone. The effect of metallicity is also shown. Bottom: Same for the stars kinematically associated with the Sirius cluster/moving group and the 300~Myr isochrone.}
}
\label{Fig:HR}
\end{figure}

We plot in HR diagrams (Fig.~3) the stars kinematically associated with the various groups (i.e. stars in the regions defined by the $V$-strip and within $1.4\sigma$ of the mean of the $U$-gaussian hereabove), together with the isochrone of the asscociated cluster (100~Myr for the Pleiades, 600~Myr for the Hyades, and 300~Myr for Sirius). Let us stress again that these HR diagrams mix stars from the background and from the moving groups/clusters. Because we perform  an analysis in  parallax space hereafter, we chose to plot the (biased) stellar absolute magnitude $M_{V_T}$ obtained directly from the inversion of the Hipparcos parallax and from the Hipparcos  magnitude $Hp$ and the $Hp - V_T$ color index, instead of the unbiased absolute magnitude of Famaey et al.~(2005). The absorption correction $A_V$ taken from Famaey et al. (2005) has however been applied.   Note that those absolute magnitudes  plotted are thus subject to the Lutz-Kelker bias but that, again, this is not a problem since the  comparison between  isochrone and  star locations will be made  in the parallax  space.  Finally,  the $V-I_C$ index, taken as well from Famaey et al. (2005), is derived from the $Hp-V_T$ index using the colour transformation from Platais et al. (2003). Isochrones  from   the  basic  grid   (labelled  'c')  of   Schaller et al. (1992) (metallicity $Z = 0.02$, standard mass loss, core overshoot, OPAL opacities) have been used, as reprocessed by  Lejeune \& Schaerer (2001) to provide the photometric indices from the Johnson $UBV$ and Cousins $RI$ bands. 

We then computed for each star in the kinematical samples the parallax  $\varpi_{\rm iso}$ expected  if the star would lie  along the associated cluster isochrone.  For the region  where RGB  and AGB overlap,  there are more  stars in  the He-clump  or on  the RGB,  since the lifetime  is longer  in these  two phases  than on  the AGB.  Therefore, the expected  parallax   is  computed  from   those  parts  of   the  isochrones corresponding  to   the  He-clump   and  RGB. Note that, since the isochrones of Lejeune \& Schaerer (2001) do not cover the whole range of observed $V-I_C$ colors, the number of stars to which an ``isochrone parallax" could be assigned is slightly smaller than the total number of stars in the region of velocity space considered.

The   normalized  difference between the ``isochrone parallax" $\varpi_{\rm  iso}$ and the measured Hipparcos parallax $\varpi_{\rm  HIP}$ 
\begin{equation}
\delta = \frac{\varpi_{\rm  HIP} -  \varpi_{\rm  iso}}{\sigma_{\varpi_{\rm HIP}}}
\end{equation}
should then be a random variable following a gaussian of mean 0 and standard dispersion 1 for a sample of coeval stars falling on the  isochrone. Knowing the fraction of stars $f_s$ that make up the overdensity superimposed on the background in each of the regions considered in velocity space (46\% in the Pleiades zone, 52\% in the Hyades zone, and 35\% in the Sirius zone), we can overplot the corresponding gaussian histogram on the total histogram of $\delta$ in the region: this gaussian histogram will be equal to
\begin{equation}
\tilde{N} = N \times \Delta_\delta \times f_s \times {\cal N}(0,1),
\end{equation}
where $N$ is the total number of stars to which an isochrone parallax has been assigned in the region considered and $\Delta_\delta$ is the width of the bins of the histogram.

Then, if a moving group is solely associated with the evaporation of its associated cluster, the subtraction of $\tilde{N}$ from the total histogram of $\delta$ in the velocity-space region should yield the typical distribution of $\delta$ for the background population. If this is not the case, the moving group will be proven to be, at least partially, linked with a dynamical (resonant) mechanism.

\subsubsection{The Pleiades moving group}

\begin{figure}
\centering
\includegraphics[width=7cm]{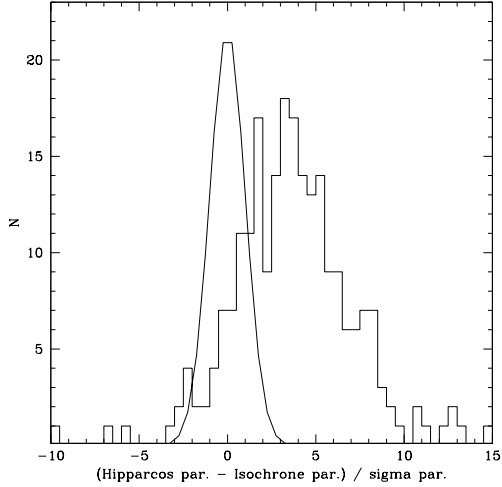}
\caption{Histogram of $\delta$ (see Eq.~7) for the Pleiades moving group, together with the gaussian $\tilde{N}$ (see Eq.~8) expected for the Pleiades moving group if all stars in the group were on the 100~Myr isochrone. The subtraction of $\tilde{N}$ from the observed histogram is negative (significantly with respect to Poissonian errors) at $\delta=0$.
}
\label{Fig:Pleiades}
\end{figure}

The histogram of $\delta$ for the 100~Myr isochrone has been plotted on Fig.~4, together with the gaussian histogram $\tilde{N}$ corresponding to the hypothesis that all stars of the Pleiades moving group (i.e. 46\% of the stars in the Pleiades velocity-zone) are falling on the 100~Myr isochrone. Very clearly, the actual amount of stars really present in the $\delta$-interval covered by the gaussian is much too low as compared to $\tilde{N}$ to be compatible with the hypothesis that 46\% of the stars are coeval at 100~Myr. If one would subtract $\tilde{N}$ to the actual histogram to find the background distribution in parallax space, one would find that the background density distribution is highly negative close to the 100~Myr isochrone, which would of course not make any sense. 

This is thus a robust proof that the Pleiades moving group is not associated with the evaporation of the Pleiades cluster. Note that the excess of stars in the $\delta$ histogram compared to the expectation for an evaporated cluster is more prominent for positive values of $\delta$: these stars are closer (and thus intrinsically fainter) than what they should be to fall on the 100~Myr isochrone, meaning that correcting for the Malmquist bias (i.e. the fact that intrinsically faint objects are missing from such a magnitude-limited sample) would make the problem even worse. Actually, a qualitative look at Fig.~3 reveals that 15 stars are clearly falling close to the 100~Myr isochrone, and are thus likely members of the evaporated Pleiades cluster.

\subsubsection{The Hyades moving group}

\begin{figure}
\centering
\includegraphics[width=7cm]{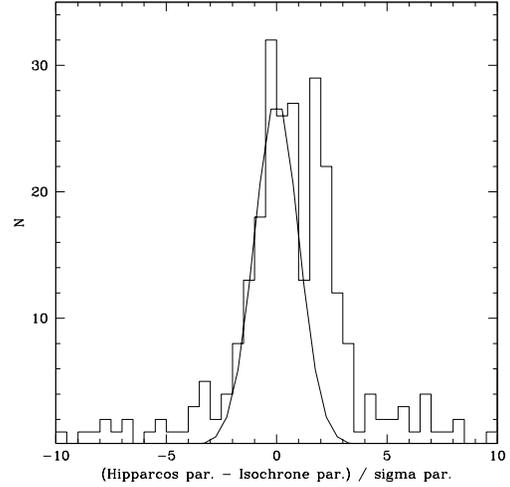}
\caption{Same as Fig.~4 for the Hyades moving group and the 600~Myr isochrone.
}
\label{Fig:Hyades}
\end{figure}

\begin{figure}
\centering
\includegraphics[width=8cm]{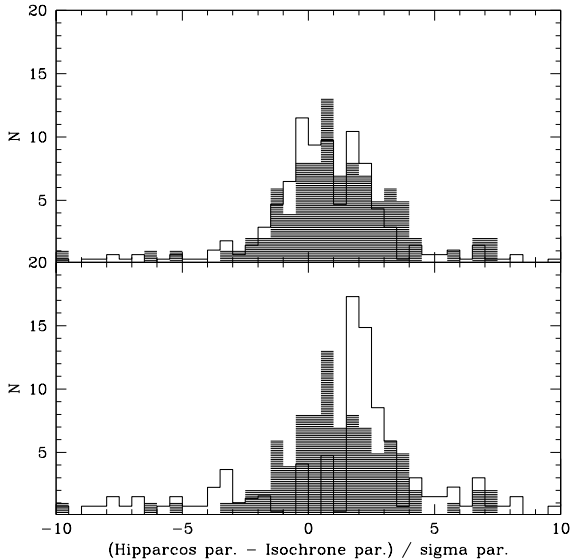}
\caption{The shaded area is the histogram of $\delta$ for the 600~Myr isochrone in the comparison box $U=[21,35]$, $V=[-24,-12]$, compared with (top) the histogram of $\delta$ in the Hyades box (background+moving group), and (bottom) the histogram of $\delta$ expected for the background if all stars of the Hyades overdensity were coeval (600~Myr), after subtracting the $\tilde{N}$ histogram of Fig.~5. All histograms are normalized to the number of stars in the comparison box.
}
\label{Fig:compa_Hyades}
\end{figure}

The histogram of $\delta$ for the 600~Myr isochrone has been plotted on Fig.~5, together with the gaussian histogram $\tilde{N}$ corresponding to the hypothesis that all stars of the Hyades moving group (i.e. 52\% of the stars in the Hyades velocity-zone) are falling on the 600~Myr isochrone. Very clearly, subtracting $\tilde{N}$ from the observed histogram would imply that the background population on which the Hyades group is superimposed has a gap in parallax space close to the 600~Myr isochrone. 

To check whether this is the case, we computed the $\delta$-histogram in a comparison box symmetrical to the Hyades with respect to $U=-2$, i.e. $U=[21,35]$, $V=[-24,-12]$. This histogram (grey-shaded in Fig.~6) should be representative of the $\delta$-distribution of the background population in the Hyades velocity-zone. We see in Fig.~6 that it is in {\it disagreement} with the $\delta$-distribution in the Hyades velocity-zone after subtracting $\tilde{N}$ (Fig.~6 bottom; normalized to the number of stars in the comparison box), but {\it compatible} with the full $\delta$-distribution in the Hyades velocity zone (Fig.~6 top). We tested the influence on our result of the metallicity adopted for the isochrone (see Fig.~3), and of the actual age assigned to the Hyades cluster (by trying 800~Myr instead of 600), and found the same disagreement. We thus conclude that it is impossible that the Hyades moving group, representing 52\% of the stars in the Hyades velocity-zone, is entirely made of stars evaporated from the Hyades cluster.

\subsubsection{The Sirius moving group}

\begin{figure}
\centering
\includegraphics[width=7cm]{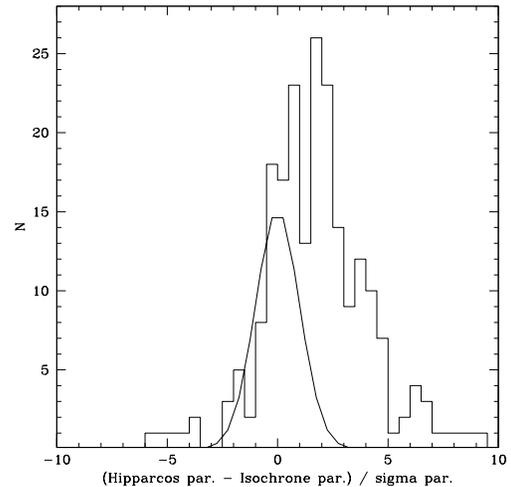}
\caption{Same as Figs.~4 and 5 for the Sirius moving group and the 300~Myr isochrone.
}
\label{Fig:Sirius}
\end{figure}

\begin{figure}
\centering
\includegraphics[width=8cm]{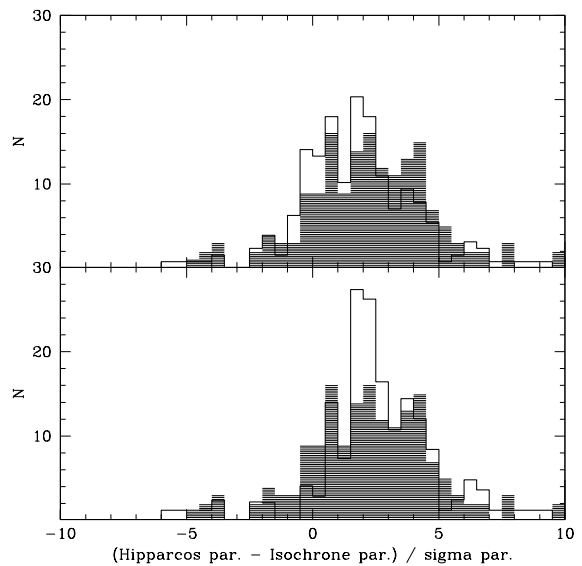}
\caption{The shaded area is the histogram of $\delta$ for the 300~Myr isochrone in the comparison box $U=[-16,2]$, $V=[-5,8]$, compared with (top) the histogram of $\delta$ in the Sirius box (background+moving group), and (bottom) the histogram of $\delta$ expected for the background if all stars of the Sirius overdensity were coeval (300~Myr), after subtracting the $\tilde{N}$ histogram of Fig.~7. All histograms are normalized to the number of stars in the comparison box.
}
\label{Fig:compa_Sirius}
\end{figure}

The histogram of $\delta$ for the 300~Myr isochrone has been plotted on Fig.~7, together with the gaussian histogram $\tilde{N}$ corresponding to the hypothesis that all stars of the Sirius moving group (i.e. 35\% of the stars in the Sirius velocity-zone) are falling on the 300~Myr isochrone. 

We then computed (grey-shaded in Fig.~8) the $\delta$-histogram in a comparison box symmetrical to Sirius with respect to $U=-2$, i.e. $U=[-16,-2]$, $V=[-5,8]$. This histogram should be representative of the $\delta$-distribution of the background population in the Sirius velocity-zone. We compared it to both the full $\delta$-distribution in the Sirius velocity zone, and to the $\delta$-distribution in the Sirius velocity-zone after subtracting $\tilde{N}$. In each case the histogram was normalized to the number of stars in the comparison box. Once again, while the observed histogram of $\delta$ in the comparison box is compatible with the total one in the Sirius box, it is in Poissonian disagreement with the histogram obtained after subtracting $\tilde{N}$ (especially around $\delta=2$). However, the disagreement is less obvious than for the Hyades: this is because the Sirius overdensity superimposed on the background is much less prominent than the Hyades one (see Fig.~2).

\section{Conclusion and perspectives}

We have reanalyzed the data of Famaey et al.~(2005) by using a wavelet transform in order to identify the main low-velocity moving groups on scales smaller than 30~km/s, and by subsequently comparing  the location of moving group stars in parallax space to isochrones  of the  embedded clusters. 

This leads to the fascinating result that moving groups do appear in  pairs in  the  [$U,V$]-diagram (The Hyades and Pleiades groups are similar to the two distinct structures in the Hercules stream), which could provide new interesting constraints on dynamical models designed to reproduce these features (see e.g. figure 9 of De Simone et al. 2004, figure 6 of Quillen \& Minchev 2005, and figure 10 of Chakrabarty 2007) if their origin is in resonant trapping by the bar, spiral arms, or a mix of both. It would also be of high interest to check whether some differences in metallicities could be observed between the two distinct structures of the Hercules stream to check whether they could have an origin at different galactocentric radii.

Moreover, this study confirms the dynamical (resonant) nature  of the Pleiades, Hyades and Sirius moving groups (see also Famaey et al. 2007), because the fraction of stars making up each velocity-space overdensity superimposed on the background is higher than the fraction of stars compatible with the isochrone of the embedded cluster. The similarity between the parallax distribution with respect to the isochrone in the velocity-zones representative of the background and of the moving groups is in fact striking (see Figs.~6 and 8), and again demonstrates that the group stars share the same distribution as the background stars in the HR diagram. These low-velocity moving groups have thus a different nature than, e.g., the HR1614 group ($U\simeq[0,20]$, $V\simeq[-65, -45]$, see De Silva et al. 2007) which was recently proven to be an evaporated cluster.

The next check of this resonant origin of the Pleiades, Hyades and Sirius moving groups should be a detailed chemical tagging of stars kinematically associated with the groups (Freeman \& Bland-Hawthorn 2002) from their element abundance patterns. For instance, we have identified here 290 giant stars kinematically associated with the Hyades, out of which 150 are making up the Hyades overdensity in velocity space (or Hyades moving group). In the resonant scenario, we expect {\it less} than 150 giant stars in this sample to have chemical abundances in precise accordance with those of the Hyades cluster (De Silva et al. 2006), although we still expect {\it some} of them to be part of the evaporating cluster. Note that we would however expect those 150 stars to be more metal-rich on average than the background because of the likely origin of the moving group in the inner parts of the Galaxy (see e.g. Daflon \& Cunha 2004 for the galactic metallicity gradient). In the case of the Sirius moving group, since the overdensity superimposed on the background is much less prominent than the Hyades one (see Fig.~2), and since the evaporated Ursa Major cluster could still make a large part of the moving group, the real dynamical overdensity should not represent as large a part of the sample as for the Hyades in such a chemical tagging analysis. Finally, another independent check of the resonant origin of the moving groups would be a detailed study of the individual kinematics of open clusters in the galactic disk (e.g., Frinchaboy 2006): within the resonant scenario, open clusters with a wide range of ages should show the same overdensities in the [$U,V$]-diagram as the individual stars studied in this paper.

\acknowledgements{We thank Pavel Kroupa for useful discussions, which partly triggered the present study}

\end{document}